\documentclass{article}

\usepackage{amsmath}
\usepackage{amssymb}
\usepackage{graphicx}

\begin{document}

\title{holographic superconductor in AdS3 spacetime}
\author{Liancheng Wang and Feng He}
\maketitle

\begin{abstract}
We do more discussions of holographic s-wave superconductors in the background of AdS3 spacetime. We analytically compute the holographic superconductor model with the motion equation has two different characteristic root on the framework of Maxwell electrodynamics field. We study superconductor models on the framework of Born-Infeld electrodynamics field and on the framework of the St\"uckelberg form in the background of AdS3 spacetime.

\end{abstract}

\section{introduction}

The AdS/CFT correspondence\cite{Maldacena1998} indicates that a weak
coupling gravity theory in (d+1)-dimensional AdS spacetime can be
described by a strong coupling conformal field theory on the d-dimensional
boundary. This means that the AdS/CFT correspondence could help us
understand more deeply about the strong coupled gauge theories\cite{Witten1998}\cite{Gubser1998}\cite{Aharony2000}
and offer us a new means of studying the strongly interacting condensed
matter systems in which the perturbation methods are no longer valid\cite{Policastro2002}\cite{Hartnoll2007b}\cite{Buchbinder2009}.
Recently, this holographic method is also expected to give us some
insights into the nature of the pairing mechanism in the high temperature
superconductors which is beyond the scope of current theories of superconductivity.
The earliest models for holographic superconductors in the AdS black
hole spacetime are proposed in \cite{Gubser2008}\cite{Hartnoll2008}\cite{Basu2009}\cite{Herzog2009a},
where the black hole admits scalar hair at the temperature $T$ smaller
than a critical temperature $T_{c}$, but does not possess scalar
hair at higher temperatures. According to the AdS/CFT correspondence,
the appearance of a hairy AdS black hole at low temperature implies
the formation of the scalar condensation in the boundary CFT, which
makes the expectation value of charged operators undergo the U(1)
symmetry breaking and results in the occurrence of the phase transition.
Due to its potential applications to the condensed matter physics,
the properties of the holographic superconductors in the various theories
of gravity have been investigated extensively\cite{Gubser2008a}\cite{Horowitz2008}\cite{Hartnoll2009}\cite{Herzog2009}\cite{Albash2008}
\cite{Nakano2008}\cite{Horowitz2009}\cite{Gregory2009}\cite{Franco2010}\cite{Chen2010}\cite{Jing2012}\cite{Pan2011}\cite{Jing2011a}
in recent years.

AdS3 spacetime as an interesting gravity model has been studied in
many works since it proposed in \cite{Banados1992}. Some holographic
properties and superconductor phase transitions in AdS3 spacetime
background have been discussed in \cite{Maity2010}\cite{Ren2010}\cite{Liu2011}\cite{Li2012}.
But up to now, researches of holographic superconductor on AdS3 spacetime are obviously less than on other spacetimes, and researches are limited to the second phase transition based on the framework of Maxwell electrodynamics.
In this paper, we will do more discussion on holographic superconductors on the framework of Maxwell electrodynamics and on the framework of Born-Infeld electrodynamics, and give the properties of holographic superconductors on the framework of the St\"uckelberg mechanism form in AdS3 spacetime.

On the framework of Maxwell electrodynamics, reference \cite{Ren2010} and \cite{Li2012} have respectively given the numerical solution and analytical solution in the model of one-dimensional holographic superconductor when the mass of scalar field is equal to Breitenlohner-Freedman bound\footnote{Reference \cite{Li2012} has also given another case that the mass of scalar field is zero.}, i.e. $m^2=-1$, the scalar field equation has two same characteristic root  $\Delta_{\pm}=1$ and its asymptotically behaves are $\psi\sim z\ln z$ and $\psi\sim z$.
Furthermore we will discuss the general situation that the equation has two different characteristic root $\Delta_{\pm}$, and the scalar field asymptotically behaves are $\psi\sim z^{\Delta_{\pm}}$.

It is well known that the properties of holographic superconductors
depend on behaviors of the electromagnetic field coupled with the
scalar filed in the system. critical temperature and the critical exponent near the point

Different from Maxiwell electrodynamics, Born-Infeld electrodynamics\cite{Born1934}
is one of the important nonlinear electromagnetic theories, which
was introduced by Born and Infeld in 1934 to deal with the infinite
self energies for charged point particles arising in Maxwell theory.
As a candidate of many improved models of Maxwell theory, it is the only electric-magnetic duality invariance theory\cite{Gibbons1995}
which has been researched extensively\cite{Dirac1931}\cite{Hooft1974}\cite{Seiberg1994}.
In particular, the effects of Born-Infeld electrodynamics on the holographic superconductors has been studied numerically in \cite{Jing2010}\cite{Jing2011}\cite{Wang2012}\cite{G.Siopsis2012}\cite{H.B.Zeng2011}\cite{H.F.Li2011}\cite{D.Momeni2012}\cite{J.A.Hutasoit2012}\cite{Gangopadhyay2012a}\cite{Pan2012}\cite{Gangopadhyay2012}\cite{Gangopadhyay2012b}.
In this paper, we are going to investigate how the Born-Infeld
electrodynamics affect the holographic superconductors phase transition in AdS3 spacetime background.

We will also discuss a generalization of the basic holographic
superconductor model in which the spontaneous breaking of a global
U(1) symmetry occurs via the St\"uckelberg mechanism\cite{Franco2010}\cite{Franco2010a}\cite{Aprile2010} in AdS3 spacetime background.
The generalized St\"uckelberg mechanism of symmetry breaking can describe
a wider class of phase transitions including the first order phase
transition and the second order phase transition.
We are going to investigate how the St\"uckelberg form affect the critical temperature, the order of phase transitions and the critical exponent near the phase transition point in AdS3 spacetime background.

This paper is organized as follows.
First we discuss the holographic superconductor model with two different characteristic root on the frame of Maxwell electrodynamics in AdS3 spacetime in section 2.
Then an analytically study of the scalar condensation and the phase transitions of holographic superconductor on the frame of Born-Infeld electrodynamics in AdS3 spacetime is given.
In Sect.4, we calculate a general class of the holographic superconductor model on the frame of the St\"uckelberg form in AdS3 spacetime.
The last section is the conclusion which contains our main results.

\section{AdS3 superconductors on Maxwell electrodynamics}

On the framework of Maxwell electrodynamics, reference \cite{Ren2010} and \cite{Li2012} have respectively given the numerical solution and analytical solution with the motion equation has two same characteristic root in the model of one-dimensional holographic superconductor. However, in this section, we will give more discussion on AdS3 holographic superconductors analytically that the motion equation has two different characteristic root on the framework of Maxwell electrodynamics.

The line element of the AdS3 black hole can be written as

\begin{equation}
ds^{2}=-\frac{1}{z^{2}}f(z)dt^{2}+\frac{dz^{2}}{f(z)}+dx^{2},\label{linez}
\end{equation}
where $f(z)=-M+r_{+}^{2}/z^{2}l^{2}$. Its Hawking temperature is
\begin{equation}
T=\frac{r_{+}}{2\pi}.\label{temperature}
\end{equation}

The action for a Maxwell electromagnetic field coupling with a charged
scalar field in AdS3 spacetime reads

\[
\mathcal{L}=-\frac{1}{4}F^{\mu\nu}F_{\mu\nu}-|\nabla_{u}\psi-iA_{\mu}\psi|^{2}-V(|\psi|).
\]
The equations of the motions are described by
\begin{eqnarray}
\phi''+\frac{1}{z}\phi'-\frac{2\psi^{2}}{z^{2}(1-z^{2})}\phi=0,\label{equationszphi}\\
z\psi''-\frac{1+z^{2}}{1-z^{2}}\psi'+\left[\frac{z\phi^{2}}{r_{+}^{2}(1-z^{2})^{2}}-\frac{m^{2}}{z(1-z^{2})}\right]\psi=0.\label{equationszpsi}
\end{eqnarray}

Unlike Ref.\cite{Li2012}, here we will discuss the situation that the motion equation has two different characteristic root $\Delta_\pm$, and at the spatial infinity, the matter fields have the form
\begin{eqnarray}
\psi=\psi^{\pm}z^{\Delta_{\pm}},\label{boundarypsi}\\
\phi=\mu\ln z+\rho,\label{boundaryphi}
\end{eqnarray}
where $\Delta_{\pm}=1\pm\sqrt{1+m^{2}}$. According to the AdS/CFT
correspondence, the dual relation of scalar operators $\langle\mathcal{O}_{\Delta_{\pm}}\rangle$
and scalar field $\psi^{\pm}$ is
\begin{equation}
\langle\mathcal{O}_{\Delta_{\pm}}\rangle\sim r_{+}^{\Delta_{\pm}}\psi^{\pm}.\label{operators}
\end{equation}

At the critical temperature $T_{c}$, $\psi=0$, so Eq.(\ref{equationszphi})
reduces to
\begin{equation}
\phi''+\frac{1}{z}\phi'=0.\label{tcphiequ}
\end{equation}
Let us set $\phi(z)=\lambda r_{+c}\ln z,~\lambda=\mu/r_{+c}$ .
Near the boundary, we introduce a new function $F(z)$ which satisfies
$\psi(z)=z^{\Delta}F(z)\langle\mathcal{O}_{\Delta}\rangle/r_{+}^{\Delta}$,
where $F(0)=1$.  At $T\rightarrow T_{c}$, the field equation of $\psi$ becomes
\begin{equation}
-F''+\frac{1}{z}\left(\frac{1+z^{2}}{1-z^{2}}-2\Delta\right)F'+\frac{\Delta^{2}}{1-z^{2}}F=\frac{\lambda^{2}\ln^{2}z}{(1-z^{2})^{2}}F\label{lambdeequlast}
\end{equation}
to be sloved subject to the boundary condition $F'(0)=0$.

According to the Sturm-Liouville eigenvalue problem, the minimum eigenvalue
$\lambda$ is
\begin{equation}
\lambda^{2}=\frac{\int_{0}^{1}dz~z^{-1+2\Delta}[(1-z^{2})F'(z)^{2}+\Delta^{2}F(z)^{2}]}{\int_{0}^{1}dz~z^{-1+2\Delta}\frac{\ln^{2}z}{1-z^{2}}F(z)^{2}}.\label{lambdaissimp}
\end{equation}
We use $F(z)$ as the following trial function
\begin{equation}
F=F_{\alpha}(z)\equiv1-\alpha z^{2}.\label{trialF}
\end{equation}
If $\Delta=1/2$, we have

\begin{equation}
\lambda_{\Delta=1/2}^{2}=\frac{\frac{1}{4}-\frac{\alpha}{6}+\frac{7\alpha^{2}}{12}}{4\alpha-\frac{56\alpha^{2}}{27}+\frac{7}{4}(\alpha-1)^{2}\zeta(3)}.\label{delta12}
\end{equation}
When $\alpha\approx0.12$, it has the minimum $\lambda^{2}\approx0.115$.
If $\Delta=3/2$, we have

\begin{equation}
\lambda_{\Delta=3/2}^{2}=\frac{\frac{3}{4}-\frac{9\alpha}{10}+\frac{11\alpha^{2}}{20}}{-2-\frac{2\alpha(-7000+3527\alpha)}{3375}+\frac{7}{4}(\alpha-1)^{2}\zeta(3)}.\label{delta32}
\end{equation}
When $\alpha\approx0.60$, it has the minimum $\lambda^{2}\approx5.586$.

And the critical temperature can be deduced\cite{Li2012}
\begin{equation}
T_{c}=\frac{r_{+c}}{2\pi}=\frac{1}{2\pi}\frac{\mu}{\lambda}.\label{tc1/2}
\end{equation}

Away from (but close to) the critical temperature, the field equation
(\ref{equationszphi}) of $\phi$ is
\begin{equation}
\phi''+\frac{1}{z}\phi'-\frac{2\langle\mathcal{O}_{\Delta}\rangle^{2}}{r_{+}^{2\Delta}}\frac{z^{2(\Delta-1)}F^{2}(z)}{1-z^{2}}\phi=0.\label{closephi}
\end{equation}
Because the parameter $\langle\mathcal{O}_{\Delta}\rangle^{2}/r_{+}^{2\Delta}$
is small, we can expand $\phi(z)$ in the parameter $\langle\mathcal{O}_{\Delta}\rangle^{2}/r_{+}^{2\Delta}$
\begin{equation}
\frac{\phi}{r_{+}}=\lambda\ln z+\frac{\langle\mathcal{O}_{\Delta}\rangle^{2}}{r_{+}^{2\Delta}}\chi(z)+\cdots.\label{expando}
\end{equation}
Substituting the above formulation into Eq.(\ref{closephi}), the
equation of $\phi$ is translated into the equation of $\chi$
\begin{equation}
\chi''+\frac{1}{z}\chi'=\lambda\frac{z^{2(\Delta-1)}F^{2}(z)}{1-z^{2}}\ln z,\label{chiequ}
\end{equation}
where the boundary condition becomes $\chi(1)=0$.

According to \cite{Siopsis2010}\cite{Li2012}, its solution can be
written as
\begin{equation}
z\chi'_{1}(0)=\lambda\mathcal{C},\qquad\mathcal{C}=\int_{0}^{1}dz\frac{z^{2\Delta-1}F^{2}(z)}{1-z^{2}}\ln z,\label{ched}
\end{equation}
then we have the expression
\begin{equation}
\frac{\mu}{r_{+}}=\lambda\left(1+\frac{\mathcal{C}\langle\mathcal{O}_{\Delta}\rangle^{2}}{r_{+}^{2\Delta}}+\cdots\right)\label{muexpand}
\end{equation}
and the operation $\langle\mathcal{O}_{\Delta}\rangle$ near the critical
temperature
\begin{equation}
\langle\mathcal{O}_{\Delta}\rangle\approx\gamma T_{c}^{\Delta}\left(1-\frac{T}{T_{c}}\right)^{\frac{1}{2}},\qquad\gamma=\frac{(2\pi)^{\Delta}}{\sqrt{\mathcal{C}}}.\label{Oexpand}
\end{equation}

\section{AdS3 superconduction on Born-Infeld electrodynamics}

In this section we will give the holographic superconductor on
the frame of Born-Infeld electromagnetic field in AdS3 spacetime.

\subsection{superconduction phase in Born-Infeld electrodynamics}

We rewriter the three-dimensional AdS metric Eq.(\ref{linez}) in the form\cite{Banados1992}
\begin{equation}
ds^{2}=-f(r)dt^{2}+f(r)^{-1}dr^{2}+r^{2}dx^{2}.\label{btzform}
\end{equation}
The Hawking temperature of black hole is
\begin{equation}
T_{H}=\frac{1}{2\pi}r_{+},\label{Hawking temperature}
\end{equation}
where $r_{+}$ is the event horizon of the black hole.

We consider a Born-Infeld field and a charged complex scalar field
coupled via the action
\begin{equation}
S=\int d^{4}x\sqrt{-g}\left(\mathcal{L}_{BI}-\left|\nabla_{\mu}\psi-iqA_{\mu}\psi\right|^{2}-m^{2}\left|\psi\right|^{2}\right)\label{L}
\end{equation}
with
\begin{equation}
\mathcal{L}_{BI}=\frac{1}{b}\left(1-\sqrt{1+\frac{bF}{2}}\right),\label{L-bi}
\end{equation}
where $F\equiv F_{\mu\nu}F^{\mu\nu}$.

Taking the ansatz that $\psi=\psi(r)$ and $A_{\mu}$ has only the
time component $A_{t}=\phi$, we can get the equations of motion for
the scalar field $\psi$ and gauge field $\phi$ in the form
\begin{eqnarray}
\phi''+\frac{1}{r}\phi'(1-b\phi'^{2})-\frac{2\psi^{2}}{f}\phi(1-b\phi'^{2})^{\frac{3}{2}}=0,\label{phi-equ-r}\\
\psi''+\left(\frac{f'}{f}+\frac{1}{r}\right)\psi'+\frac{\phi^{2}}{f^{2}}\psi-\frac{m^{2}}{f}\psi=0.\label{psi-equ-r}
\end{eqnarray}

Near the boundary $r\rightarrow\infty$, the matter fields have the asymptotic behaviors
\begin{eqnarray}
 &  & \psi=\frac{\psi^{\pm}}{r^{\triangle_{\pm}}},\label{psi-bound}\\
 &  & \phi=-\mu\ln r+\rho,\label{phi-bound}
\end{eqnarray}
where $\mu$ and $\rho$ are interpreted as the chemical potential and charge density in the dual field theory respectively. From the dual field theory, the dual relation of scale operate $\mathcal{O}$ and field $\psi_{\pm}$ can be written as
\begin{equation}
\langle\mathcal{O}_{\pm}\rangle\sim\psi_{\pm},\label{opsi-r}
\end{equation}

In Figure 1 we present the condensation of the scalar operators $\langle{\cal O}_{+}\rangle$
as a function of temperature with various correction terms $b$. We
find that the condensation gap increases with the Born-Infeld scale
parameter $b$, which means that the scalar hair is harder to be formed
than the usual Maxwell field. This behavior is reminiscent of that
seen for the 4-dimension holographic superconductors with Born-Infeld
electrodynamics\cite{Jing2010}, where the higher Born-Infeld corrections make condensation
harder.

\begin{figure}[htbp]
\centering \includegraphics[scale=1.5]{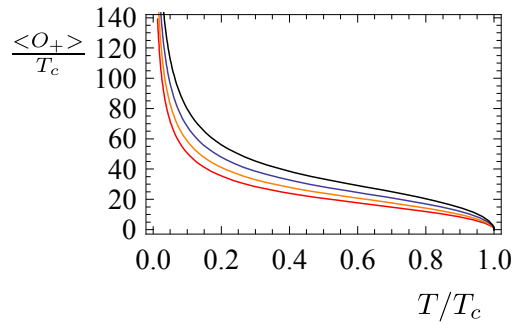}
\caption{(Color online) The condensate $\langle\mathcal{O}_{i}\rangle$ as
a function of temperature $T$ with different $b$. The four lines
correspond to increasing $b$, i.e., $0$ (red), $0.01$ (orange),
$0.02$ (blue), $0.03$ (black) from bottom to top as $\triangle=1/2$.}

\label{figure1}
\end{figure}

\subsection{analytical understanding of critical temperature}

Here we will apply the Sturm-Liouville method\cite{Siopsis2010} to
analytically investigate the properties of holographic superconductor
phase transition with Born-Infeld electromagnetic field.

Introducing a new variable $z=r_{+}/r$, we can rewrite Eq.(\ref{phi-equ-r})
and Eq.(\ref{psi-equ-r}) into
\begin{eqnarray}
\phi'(z)+\frac{bz^{4}\phi'(z)^{3}}{r_{+}^{2}}+\frac{2\phi(z)\psi(z)^{2}\left[1-\frac{bz^{4}\phi'(z)^{2}}{r_{+}^{2}}\right]^{\frac{3}{2}}}{-z+z^{3}}+z\phi''(z)=0,\label{phi-equ-z}\\
\left[\frac{m^{2}}{z\left(-1+z^{2}\right)}+\frac{z\phi(z)^{2}}{r_{+}^{2}\left(-1+z^{2}\right)^{2}}\right]\psi(z)+\frac{\left(1+z^{2}\right)\psi'(z)}{-1+z^{2}}+z\psi''(z)=0.\label{psi-equ-z}
\end{eqnarray}

The asymptotic boundary conditions for the scalar potential $\phi(z)$ and the scalar field $\psi(z)$ turn
out to be
\begin{eqnarray}
 &  & \psi=\psi^{\pm}z^{\triangle_{\pm}},\label{psi-bound-z}\\
 &  & \phi=\mu\ln z+\rho,\label{phi-bound-z}
\end{eqnarray}
and relation Eq.(\ref{opsi-r}) becomes
\begin{equation}
\langle\mathcal{O}_{\Delta_{\pm}}\rangle\sim r_{+}^{\Delta_{\pm}}\psi^{\pm}.
\end{equation}

With the above set up in place, we are now in a position to investigate
the relation between the critical temperature and the charge density.
At the critical temperature $T_{c}$, $\psi=0$, so the Eq.(\ref{phi-equ-z})
reduces to
\begin{equation}
\phi''(z)+\frac{bz^{3}\phi'(z)^{3}}{r_{+}^{2}}+\frac{\phi'(z)}{z}=0.\label{phi-equ-tc}
\end{equation}
Letting $\phi'(z)=1/\sqrt{\zeta}$, we have
\begin{equation}
-\frac{1}{2}\zeta'(z)+\frac{bz^{3}}{r_{+}^{2}}+\frac{\zeta}{z}=0.\label{zeta-equ}
\end{equation}
It is easy to obtain the solution of the above equation
\begin{equation}
\zeta(z)=\frac{bz^{4}}{r^{2}}+z^{2}c_{1},\label{zeta-solve}
\end{equation}
then we can obtain the solution of $\phi(z)$ with the coefficient
$c_{1}$ and $c_{2}$
\begin{equation}
\phi(z)=\frac{\ln z-\ln(c_{1}+\sqrt{c_{1}}\sqrt{c_{1}+bz^{2}})}{\sqrt{c_{1}}}+c_{2}.
\end{equation}
According to the boundary condition Eq.(\ref{phi-bound-z}) and the
horizon condition $\phi(z=1)=0$, we can solve $c_{1}$ and $c_{2}$
\begin{equation}
c_{1}=\frac{1}{\mu^{2}},~~~~c_{2}=\mu\ln\left(\frac{1}{\mu^{2}}+\frac{1}{\mu}\sqrt{b+\frac{1}{\mu^{2}}}\right).\label{c1c2}
\end{equation}
With the aid of the relation $\lambda=\mu/r_{+c}$, at last we have
the result
\begin{equation}
\phi(z)=r_{+c}\lambda\left[\ln z-\ln(1+\sqrt{1+bz^{2}\lambda^{2}})+\ln(1+\sqrt{1+b\lambda^{2}})\right].
\end{equation}
It can be expanded into a simply form
\begin{equation}
\phi(z)=r_{+c}\lambda\left(-\frac{1}{4}bz^{2}\lambda^{2}+\ln z+\frac{b\lambda^{2}}{4}\right).
\end{equation}
Using the above expansion, we find as $T\rightarrow T_{c}$ the field
equation Eq.(\ref{psi-equ-z}) of $\psi$ approaches the limit
\begin{equation}
-\psi''+\frac{1+z^{2}}{z(1-z^{2})}\psi'+\frac{m^{2}}{z^{2}(1-z^{2})}\psi=\frac{\lambda^{2}\left(\ln z-\frac{1}{4}bz^{2}\lambda^{2}+\frac{b\lambda^{2}}{4}\right)^{2}}{(1-z^{2})^{2}}\psi.\label{psi-tc-equ}
\end{equation}

Near the boundary, we introduce a new function $F(z)$ which satisfies
\begin{equation}
\psi(z)=\frac{\langle\mathcal{O}_{\Delta}\rangle}{r_{+}^{\Delta}}z^{\Delta}F(z),
\end{equation}
where $F(0)=1$. So from Eq.(\ref{psi-tc-equ}), the equation of motion
for $F(z)$ is
\begin{equation}
-F''+\frac{1}{z}\left(\frac{1+z^{2}}{1-z^{2}}-2\Delta\right)F'+\frac{\Delta^{2}}{1-z^{2}}F=\frac{\lambda^{2}\left[\ln^{2}z-\frac{1}{2}b\lambda^{2}\ln z(z^{2}-1)\right]}{(1-z^{2})^{2}}F,
\end{equation}
to be solved subject to the boundary condition $F'(0)=0$.

According to the Sturm-Liouville eigenvalue problem\cite{Siopsis2010},
we obtain the expression which will be used to estimate the minimum
eigenvalue of $\lambda$
\begin{equation}
\lambda^{2}=\frac{\int_{0}^{1}dz~z^{-1+2\Delta}[(1-z^{2})F'(z)^{2}+\Delta^{2}F(z)^{2}]}{\int_{0}^{1}dz~z^{-1+2\Delta}\frac{\ln^{2}z-\frac{1}{2}b\lambda^{2}\ln z(z^{2}-1)}{1-z^{2}}F(z)^{2}}.
\end{equation}
To estimate it, we use $F(z)$ as the following trial function
\[
F=F_{\alpha}(z)\equiv1-\alpha z^{2},
\]
which satisfies the conditions $F(0)=1$ and $F'(0)=0$.

When $\Delta=1/2$ , for $b=0$, we have
\[
\lambda_{\alpha}^{2}=\frac{27-18\alpha+63\alpha^{2}}{189\zeta(3)+[432-378\zeta(3)]\alpha+[189\zeta(3)-224]\alpha^{2}},
\]
and when $\alpha\approx0.123209$, it reaches its minimum $\lambda^{2}\approx\lambda_{0.123209}^{2}\approx0.114659$,
which is agree with the exact value $\lambda^{2}=0.113939$. The critical
temperature is
\begin{equation}
T_{c}=\frac{r_{+c}}{2\pi}=\frac{1}{2\pi}\frac{\mu}{\lambda}.\label{tc-lamda0}
\end{equation}
From Eq.(\ref{tc-lamda0}) we can obtain $T_{c}\approx0.47002\mu$,
which is agree with the exact value $T_{c}=0.471503\mu$.

When $b=0.1$, we obtain
\[
\lambda_{\alpha}^{2}=\frac{\frac{1}{4}-\frac{\alpha}{6}+\frac{7\alpha^{2}}{12}}{2.09787-0.205925\alpha+0.0292962\alpha^{2}},
\]
and when $\alpha\approx0.123276$, it reaches its minimum $\lambda^{2}\approx\lambda_{0.123276}^{2}\approx0.114967$,
which is agree with the exact value $\lambda^{2}=0.114239$. From
Eq.(\ref{tc-lamda0}) we can obtain $T_{c}\approx0.469389\mu$, which
is agree with the exact value $T_{c}=0.470884\mu$.

When $b=0.2$, we obtain
\[
\lambda_{\alpha}^{2}=\frac{\frac{1}{4}-\frac{\alpha}{6}+\frac{7\alpha^{2}}{12}}{2.09213-0.204651\alpha+0.0290669\alpha^{2}},
\]
and when $\alpha\approx0.123344$, it reaches its minimum $\lambda^{2}\approx\lambda_{0.123344}^{2}\approx0.115278$,
which is agree with the exact value $\lambda^{2}=0.114539$. From
Eq.(\ref{tc-lamda0}) we can obtain $T_{c}\approx0.468757\mu$, which
is agree with the exact value $T_{c}=0.470267\mu$.

When $b=0.3$, we obtain
\[
\lambda_{\alpha}^{2}=\frac{\frac{1}{4}-\frac{\alpha}{6}+\frac{7\alpha^{2}}{12}}{2.0864-0.203377\alpha+0.0288376\alpha^{2}},
\]
and when $\alpha\approx0.123412$, it reaches its minimum $\lambda^{2}\approx\lambda_{0.123412}^{2}\approx0.11559$,
which is agree with the exact value $\lambda^{2}=0.114838$. From
Eq.(\ref{tc-lamda0}) we can obtain $T_{c}\approx0.468124\mu$, which
is agree with the exact value $T_{c}=0.469652\mu$.

\subsection{critical exponent and condensation values}

Away from(but close to) the critical temperature, the field equation
Eq.(\ref{phi-equ-z}) of $\phi$ is
\begin{equation}
\phi''(z)+\frac{\phi'(z)}{z}+\frac{bz^{3}\phi'(z)^{3}}{r_{+}^{2}}-\frac{2\langle\mathcal{O}_{\Delta}\rangle^{2}}{r_{+}^{2\Delta}}\frac{z^{2(\Delta-1)}F^{2}(z)}{1-z^{2}}\phi(z)\left[1-\frac{3bz^{4}}{2r_{+}^{2}}\phi'(z)^{2}\right]=0.\label{phi-close-equ}
\end{equation}
Because the parameter $\langle\mathcal{O}_{\Delta}\rangle^{2}/r_{+}^{2\Delta}$
is small, we can expand $\phi(z)$ on the parameter $\langle\mathcal{O}_{\Delta}\rangle^{2}/r_{+}^{2\Delta}$
\begin{equation}
\frac{\phi}{r_{+}}=\lambda\left(\ln z-\frac{1}{4}bz^{2}\lambda^{2}+\frac{b\lambda^{2}}{4}\right)+\frac{\langle\mathcal{O}_{\Delta}\rangle^{2}}{r_{+}^{2\Delta}}\chi(z)+\cdots.\label{phi-o}
\end{equation}
Substituting the above formulation into Eq.(\ref{phi-close-equ}),
we translate the equation of $\phi$ into the equation of $\chi$
\[
\chi''+\frac{1}{z}\chi'+3b\lambda^{2}z\chi'=2\lambda\frac{z^{2(\Delta-1)}F^{2}(z)}{1-z^{2}}\left[\ln z-\frac{1}{4}b\lambda^{2}(z^{2}+6z^{2}\ln z-1)\right],
\]
where the boundary condition becomes $\chi(1)=0$.

Multiplying the two sides of this equation by $ze^{3bz^{2}\lambda^{2}/2}$,
we have
\begin{equation}
\frac{d}{dz}\left(ze^{3bz^{2}\lambda^{2}/2}\chi'\right)=ze^{3bz^{2}\lambda^{2}/2}\frac{2\lambda z^{2(\Delta-1)}F^{2}(z)}{1-z^{2}}\left[\ln z-\frac{1}{4}b\lambda^{2}(z^{2}+6z^{2}\ln z-1)\right].\label{dx-equ}
\end{equation}
The variable $\chi(z)$ in Eq.(\ref{phi-o}) can be expanded at $z=0$

\begin{equation}
\frac{\mu\ln z+\rho}{r_{+}}=\lambda\left(\ln z-\frac{1}{4}bz^{2}\lambda^{2}+\frac{b\lambda^{2}}{4}\right)+\frac{\langle\mathcal{O}_{\Delta}\rangle^{2}}{r_{+}^{2\Delta}}[\chi(0)+z\chi'(0)\cdots].\label{co-expand}
\end{equation}

Integrating both sides of Eq.(\ref{dx-equ}) from $z=0$ to $z=1$,
we have

\begin{equation}
z\chi'(0)=-\lambda\mathcal{C},\label{eq:cvalue}
\end{equation}
where

\begin{eqnarray}
\mathcal{C} & = & \int_{0}^{1}dz~e^{3bz^{2}\lambda^{2}/2}\frac{2z^{2\Delta-1}F^{2}(z)}{1-z^{2}}\left[\ln z-\frac{1}{4}b\lambda^{2}(z^{2}+6z^{2}\ln z-1)\right]\nonumber \\
 & = & \int_{0}^{1}dz~\frac{2z^{2\Delta-1}F^{2}(z)}{1-z^{2}}\left(\ln z-\frac{1}{4}bz^{2}\lambda^{2}+\frac{b\lambda^{2}}{4}\right),\label{cvalue}
\end{eqnarray}
in the last line we have expanded $\mathcal{C}$ about $b\lambda^{2}$.

With the aid of the above relation, differentiating the two sides
of the equation Eq.(\ref{co-expand}) and comparing the coefficient
of $z$ on both sides of the equation Eq.(\ref{co-expand}), we have
\begin{equation}
\frac{\mu}{r_{+}}=\lambda\left(1-\frac{\mathcal{C}\langle\mathcal{O}_{\Delta}\rangle^{2}}{r_{+}^{2\Delta}}+\cdots\right).\label{Oexpress}
\end{equation}
Combining Eq.(\ref{tc-lamda0}) and Eq.(\ref{Oexpress}), we can obtain
the express of operation $\langle\mathcal{O}_{\Delta}\rangle$ near
the critical temperature
\begin{equation}
\langle\mathcal{O}_{\Delta}\rangle\approx\gamma T_{c}^{\Delta}\left(1-\frac{T}{T_{c}}\right)^{\frac{1}{2}},\qquad\gamma=\frac{(2\pi)^{\Delta}}{\sqrt{\mathcal{C}}}.\label{tc-exponent}
\end{equation}

Combining Eq.(\ref{eq:cvalue}), Eq.(\ref{cvalue}) and Eq.(\ref{tc-exponent}),
we can obtain the solution of $\gamma$ as $\Delta=1/2$. The results
are summarized with various $b$ in Table 1:

\begin{table}[htbp]
\centering %
\begin{tabular}{|c|c|c|c|}
\hline
$b$  & $\alpha$ & $\gamma_{SL}$  & $\gamma_{numb}$ \\
\hline
0  & 0.123209 & 1.633  & 1.581 \\
0.01  & 0.123276 & 1.635  & 1.575 \\
0.02  & 0.123344 & 1.637  & 1.569 \\
0.03  & 0.123412 & 1.639  & 1.564 \\
\hline
\end{tabular}
\caption{A comparison of the analytical and numerical results for the parameter $\gamma$ between the critical temperature and the condensation operator for different $b$.}

\label{table}
\end{table}

\section{AdS3 superconduction in St\"uckelberg form}

In this section we will discuss the holographic superconductor on
the frame of St\"uckelberg form in AdS3 spacetime.

The generalized action containing a U(1) gauge field and the scalar
field coupled via a generalized St\"uckelberg Lagrangian reads\cite{Franco2010}\cite{Franco2010a}
\begin{equation}
\mathcal{L}=-\frac{1}{4}F_{\mu\nu}F^{\mu\nu}-\partial\psi^{2}-|\mathcal{K}(\psi)|(\partial p-A)^{2}-m^{2}|\psi|^{2},\label{actiongene}
\end{equation}
where $\mathcal{K}(\psi)$ is a general function of $\psi$
\begin{equation}
\mathcal{K}(\psi)=|\psi|^{2}+c_{\gamma}|\psi|^{\gamma}+c_{4}|\psi|^{4}.\label{kgene}
\end{equation}

Taking the ansatz of the field as $\psi=\psi(r)$ and $A=\phi(r)dt$,
we can get the equations of motion for scalar field $\psi$ and gauge
field $\phi$ in the form
\begin{eqnarray}
\phi''+\frac{1}{z}\phi'-\frac{\mathcal{K}}{z^{2}(1-z^{2})}\phi=0,\label{motiveequgene1}\\
z\psi''-\frac{1+z^{2}}{1-z^{2}}\psi'+\frac{z\phi^{2}}{2r_{+}^{2}(1-z^{2})^{2}}\mathcal{K}'-\frac{m^{2}}{z(1-z^{2})}\psi=0.\label{motiveequgene2}
\end{eqnarray}

At the critical temperature $T_{c}$, $\psi=0$, so Eq.(\ref{motiveequgene1}) reduces to
\begin{equation}
\phi''+\frac{1}{z}\phi'=0.\label{tcphiequgene}
\end{equation}
The asymptotic boundary conditions for the scalar potential $\phi$
turn out to be
\begin{equation}
\phi(z)=\lambda r_{+c}\ln z,\qquad\lambda=\frac{\mu}{r_{+c}}.\label{philambdagene}
\end{equation}

\subsection{$c_{\gamma}=0$}

With $c_{\gamma}=0$,
\footnote{Now the phase transition of holographic superconductor
is the second order phase transition, so we can discuss with analytical
method directly.}
at $T\rightarrow T_{c}$, the equation of $\psi$ becomes
\begin{equation}
-\psi''+\frac{1+z^{2}}{z(1-z^{2})}\psi'+\frac{m^{2}}{z^{2}(1-z^{2})}\psi=\frac{\lambda^{2}\ln^{2}z}{(1-z^{2})^{2}}(\psi+2c_{4}\psi^{3}).\label{phiequlambdagene}
\end{equation}
Near the boundary, we introduce a new function $F(z)$ which satisfies
\begin{equation}
\psi(z)=\frac{\langle\mathcal{O}_{\Delta}\rangle}{r_{+}^{\Delta}}z^{\Delta}F(z),\label{psifgene}
\end{equation}
where $F(0)=1$. Now Eq.(\ref{phiequlambdagene}) becomes
\begin{equation}
-F''+\frac{1}{z}\left(\frac{1+z^{2}}{1-z^{2}}-2\Delta\right)F'+\frac{\Delta^{2}}{1-z^{2}}F=\frac{\lambda^{2}\ln^{2}z}{(1-z^{2})^{2}}(F+2c_{4}F^{3})\label{lambdeequlastgene}
\end{equation}
to be sloved subject to the boundary condition $F'(0)=0$. Since our
computation is near the critical point, $F$ is small, so the term
$2c_{4}F^{3}$ can be ignored. It is the same as Eq.(\ref{lambdeequlast}),
so its discussion follows the process from Eq.(\ref{lambdaissimp})
to Eq.(\ref{tc1/2}).

Away from (but close to) the critical temperature, the scalar potential
$\phi$ equation is
\begin{equation}
\phi''+\frac{1}{z}\phi'-\left[\frac{\langle\mathcal{O}_{\Delta}\rangle^{2}}{r_{+}^{2\Delta}}\frac{z^{2\Delta}F^{2}(z)}{z^{2}(1-z^{2})}+\frac{c_{4}\langle\mathcal{O}_{\Delta}\rangle^{4}}{r_{+}^{4\Delta}}\frac{z^{4\Delta}F^{4}(z)}{z^{2}(1-z^{2})}\right]\phi=0.\label{closephigene}
\end{equation}
Because the parameter $\langle\mathcal{O}_{\Delta}\rangle^{2}/r_{+}^{2\Delta}$
is small, we can expand $\phi$ on the small parameter $\langle\mathcal{O}_{\Delta}\rangle^{2}/r_{+}^{2\Delta}$
\begin{equation}
\frac{\phi}{r_{+}}=\lambda\ln z+\frac{\langle\mathcal{O}_{\Delta}\rangle^{2}}{r_{+}^{2\Delta}}\chi(z)+\cdots.\label{expandogene}
\end{equation}
By substituting the above formulation into Eq.(\ref{closephigene}),
the equation of $\phi$ is translated into the equation of $\chi$
\begin{equation}
\chi''+\frac{1}{z}\chi'=\frac{\lambda\ln z}{z^{2}(1-z^{2})}\left[z^{2\Delta}F^{2}(z)+\frac{c_{4}\mathcal{O}^{2}}{r_{+}^{2\Delta}}z^{4\Delta}F^{4}(z)\right],\label{chiequgene}
\end{equation}
where the boundary condition becomes $\chi(1)=0$.

Reference \cite{Siopsis2010}, the asymptotic behavior of $\phi$
at the boundary can be written as
\begin{equation}
\frac{\mu}{r_{+}}=\lambda\left(1+\frac{\mathcal{C}_{1}\langle\mathcal{O}_{\Delta}\rangle^{2}}{r_{+}^{2\Delta}}+\frac{\mathcal{C}_{2}\langle\mathcal{O}_{\Delta}\rangle^{4}}{r_{+}^{4\Delta}}+\cdots\right).\label{muexpandgene}
\end{equation}
According to Eq.(\ref{chiequgene}), we have
\[
z\chi'_{1}(0)=\lambda\left(\mathcal{C}_{1}+\mathcal{C}_{2}\frac{\langle\mathcal{O}_{\Delta}\rangle^{2}}{r_{+}^{2\Delta}}\right),
\]
where
\begin{equation}
\mathcal{C}_{1}=\int_{0}^{1}dz\ln z\frac{z^{2\Delta-1}F^{2}(z)}{1-z^{2}},\qquad\mathcal{C}_{2}=\int_{0}^{1}dz\ln z\frac{c_{4}z^{4\Delta-1}F^{4}(z)}{1-z^{2}},\label{chedgene}
\end{equation}
and we can obtain the operation $\langle\mathcal{O}_{\Delta}\rangle$
near the critical temperature
\begin{equation}
\langle\mathcal{O}_{\Delta}\rangle\propto T_{c}^{\Delta}\left(1-\frac{T}{T_{c}}\right)^{\frac{1}{2}}.\label{Oexpandgene}
\end{equation}

\subsection{$c_{\gamma}\protect\neq0$}

With $c_{\gamma}\neq0$, if $\gamma>4$, it is the same as the situation of $c_{\gamma}=0$, so we need only discuss the situation of $2<\gamma<4$. For simplellyfi, here we will only discuss the explicit model $\gamma=3$, and we will rewrite $c_{\gamma}$ as $c_{3}$.\footnote{At this situation, The phase transition of the model may be the second order phase transition or the first order phase transition, but the analytical discussion can only be use to the second phase transition, so our discussion should be limited to the situation that the phase transition is the second order.}
At $T\rightarrow T_{c}$, the equation of $\psi$ becomes
\begin{equation}
-\psi''+\frac{1+z^{2}}{z(1-z^{2})}\psi'+\frac{m^{2}}{z^{2}(1-z^{2})}\psi=\frac{\lambda^{2}\ln^{2}z}{(1-z^{2})^{2}}\left(\psi+\frac{c_{3}}{2}\psi^{2}+2c_{4}\psi^{3}\right).\label{phiequlambdagene2}
\end{equation}
Near the boundary, we also introduce a new function $F(z)$ which
satisfies
\begin{equation}
\psi(z)=\frac{\langle\mathcal{O}_{\Delta}\rangle}{r_{+}^{\Delta}}z^{\Delta}F(z),\label{psifgene2}
\end{equation}
where $F(0)=1$. Now Eq.(\ref{phiequlambdagene2}) becomes
\begin{equation}
-F''+\frac{1}{z}\left(\frac{1+z^{2}}{1-z^{2}}-2\Delta\right)F'+\frac{\Delta^{2}}{1-z^{2}}F=\frac{\lambda^{2}\ln^{2}z}{(1-z^{2})^{2}}\left(F+\frac{c_{3}}{2}F^{2}+2c_{4}F^{3}\right)\label{lambdeequlastgene2}
\end{equation}
to be sloved subject to the boundary condition $F'(0)=0$. Since our
computation is near the critical point, $F$ is small, so the term
$c_{3}F^{2}/2+2c_{4}F^{3}$ can be ignored. It is the same as Eq.(\ref{lambdeequlast}),
and its discussion follows the process from Eq.(\ref{lambdaissimp})
to Eq.(\ref{tc1/2}).

Away from (but close to) the critical temperature, the scalar potential
$\phi$ equation is
\begin{equation}
\phi''+\frac{1}{z}\phi'-\left[\frac{\langle\mathcal{O}_{\Delta}\rangle^{2}}{r_{+}^{2\Delta}}\frac{z^{2\Delta}F^{2}(z)}{z^{2}(1-z^{2})}+\frac{c_{3}\langle\mathcal{O}_{\Delta}\rangle^{3}}{r_{+}^{3\Delta}}\frac{z^{3\Delta}F^{3}(z)}{z^{2}(1-z^{2})}+\frac{c_{4}\langle\mathcal{O}_{\Delta}\rangle^{4}}{r_{+}^{4\Delta}}\frac{z^{4\Delta}F^{4}(z)}{z^{2}(1-z^{2})}\right]\phi=0.\label{closephigene2}
\end{equation}
Because the parameter $\langle\mathcal{O}_{\Delta}\rangle^{2}/r_{+}^{2\Delta}$
is small, we can expand $\phi$ on the small parameter $\langle\mathcal{O}_{\Delta}\rangle^{2}/r_{+}^{2\Delta}$
\begin{equation}
\frac{\phi}{r_{+}}=\lambda\ln z+\frac{\langle\mathcal{O}_{\Delta}\rangle^{2}}{r_{+}^{2\Delta}}\chi(z)+\cdots.\label{expandogene2}
\end{equation}
Substituting the above formulation into Eq.(\ref{closephigene2}),
the equation of $\phi$ is translated into the equation of $\chi$
\begin{equation}
\chi''+\frac{1}{z}\chi'=\frac{\lambda\ln z}{z^{2}(1-z^{2})}\left[z^{2\Delta}F^{2}(z)+\frac{c_{3}\mathcal{O}}{r_{+}^{\Delta}}z^{3\Delta}F^{3}(z)+\frac{c_{4}\mathcal{O}^{2}}{r_{+}^{2\Delta}}z^{4\Delta}F^{4}(z)\right],\label{chiequgene2}
\end{equation}
where the boundary condition becomes $\chi(1)=0$.

Reference \cite{Siopsis2010}, the asymptotic behavior of $\phi$
at the boundary can be written as
\begin{equation}
\frac{\mu}{r_{+}}=\lambda\left(1+\frac{\mathcal{C}_{2}\langle\mathcal{O}_{\Delta}\rangle^{2}}{r_{+}^{2\Delta}}+\frac{\mathcal{C}_{3}\langle\mathcal{O}_{\Delta}\rangle^{3}}{r_{+}^{3\Delta}}+\frac{\mathcal{C}_{4}\langle\mathcal{O}_{\Delta}\rangle^{4}}{r_{+}^{4\Delta}}+\cdots\right),\label{muexpandgene2}
\end{equation}
According to Eq.(\ref{chiequgene2}), we have
\begin{equation}
z\chi'_{1}(0)=\lambda\left(\mathcal{C}_{2}+\mathcal{C}_{3}\frac{\langle\mathcal{O}_{\Delta}\rangle}{r_{+}^{\Delta}}+\mathcal{C}_{4}\frac{\langle\mathcal{O}_{\Delta}\rangle^{2}}{r_{+}^{2\Delta}}\right),\label{chedgene2}
\end{equation}
where
\begin{eqnarray}
\mathcal{C}_{2} & = & \int_{0}^{1}dz\ln z\frac{z^{2\Delta-1}F^{2}(z)}{1-z^{2}},\nonumber \\
\mathcal{C}_{3} & = & \int_{0}^{1}dz\ln z\frac{c_{3}z^{3\Delta-1}F^{3}(z)}{1-z^{2}},\nonumber \\
\mathcal{C}_{4} & = & \int_{0}^{1}dz\ln z\frac{c_{4}z^{4\Delta-1}F^{4}(z)}{1-z^{2}},\label{chedgene2c}
\end{eqnarray}
and we can obtain the operation $\langle\mathcal{O}_{\Delta}\rangle$
near the critical temperature
\begin{equation}
\langle\mathcal{O}_{\Delta}\rangle\propto T_{c}^{\Delta}\left(1-\frac{T}{T_{c}}\right).\label{Oexpandgene2}
\end{equation}

\section{conclusion}

We study holographic superconductors in the Maxwell electrodynamics field, Born-Infeld electrodynamics field and St\"uckelberg form for a planar AdS3 black hole spacetime. Analytical computations are based on the Sturm-Liouville eigenvalue problem.
On the framework of Maxwell electrodynamics, we give a discussion on AdS3 holographic superconductors analytically when the motion equation has two different characteristic root.
Then we study the scalar condensation and the phase transitions of holographic superconductor models on the frame of Born-Infeld electrodynamics. In the probe limit, we obtain the relation between the critical temperature and the charge density. Apparently, the critical temperature is affected by the value of Born-Infeld coupling parameter $b$.
Finally we calculate superconductor phase transition in St\"uckelberg form. It can be found that the critical temperature and the critical exponent are affected by the value of St\"uckelberg parameter $c_{\gamma},c_{4},\gamma$.

We would like to thank Pro. Jiliang Jing for many guidances on a draft of this paper.

\end{document}